\documentclass{eptcs}

\usepackage{amsmath}

\title{Domain-Specific Languages of Mathematics:
   Presenting Mathematical Analysis Using Functional Programming}

\author{Cezar Ionescu
\institute{Chalmers Univ. of Technology}
\email{cezar@chalmers.se}
\and
Patrik Jansson
\institute{Chalmers Univ. of Technology}
\email{\quad patrikj@chalmers.se}
}

\def\titlerunning{DSLs of Mathematics}
\def\authorrunning{C. Ionescu \& P. Jansson}
\newcommand{\event}{4th International Workshop on Trends in Functional Programming in Education, TFPIE 2015}

\makeatletter
\@ifundefined{lhs2tex.lhs2tex.sty.read}%
  {\@namedef{lhs2tex.lhs2tex.sty.read}{}%
   \newcommand\SkipToFmtEnd{}%
   \newcommand\EndFmtInput{}%
   \long\def\SkipToFmtEnd#1\EndFmtInput{}%
  }\SkipToFmtEnd

\newcommand\ReadOnlyOnce[1]{\@ifundefined{#1}{\@namedef{#1}{}}\SkipToFmtEnd}
\usepackage{amstext}
\usepackage{amssymb}
\usepackage{stmaryrd}
\DeclareFontFamily{OT1}{cmtex}{}
\DeclareFontShape{OT1}{cmtex}{m}{n}
  {<5><6><7><8>cmtex8
   <9>cmtex9
   <10><10.95><12><14.4><17.28><20.74><24.88>cmtex10}{}
\DeclareFontShape{OT1}{cmtex}{m}{it}
  {<-> ssub * cmtt/m/it}{}

\DeclareFontShape{OT1}{cmtt}{bx}{n}
  {<5><6><7><8>cmtt8
   <9>cmbtt9
   <10><10.95><12><14.4><17.28><20.74><24.88>cmbtt10}{}
\DeclareFontShape{OT1}{cmtex}{bx}{n}
  {<-> ssub * cmtt/bx/n}{}

\newcommand{\Conid}[1]{\mathit{#1}}
\newcommand{\Varid}[1]{\mathit{#1}}
\newcommand{\anonymous}{\kern0.06em \vbox{\hrule\@width.5em}}
\newcommand{\plus}{\mathbin{+\!\!\!+}}

\renewcommand{\leq}{\leqslant}
\renewcommand{\geq}{\geqslant}
\usepackage{polytable}

\@ifundefined{mathindent}%
  {\newdimen\mathindent\mathindent\leftmargini}%
  {}%

\def\resethooks{%
  \global\let\SaveRestoreHook\empty
  \global\let\ColumnHook\empty}
\newcommand*{\savecolumns}[1][default]%
  {\g@addto@macro\SaveRestoreHook{\savecolumns[#1]}}
\newcommand*{\restorecolumns}[1][default]%
  {\g@addto@macro\SaveRestoreHook{\restorecolumns[#1]}}
\newcommand*{\aligncolumn}[2]%
  {\g@addto@macro\ColumnHook{\column{#1}{#2}}}

\resethooks

\newcommand{\onelinecommentchars}{\quad-{}- }
\newcommand{\commentbeginchars}{\enskip\{-}
\newcommand{\commentendchars}{-\}\enskip}

\newcommand{\visiblecomments}{%
  \let\onelinecomment=\onelinecommentchars
  \let\commentbegin=\commentbeginchars
  \let\commentend=\commentendchars}

\newcommand{\invisiblecomments}{%
  \let\onelinecomment=\empty
  \let\commentbegin=\empty
  \let\commentend=\empty}

\visiblecomments

\newlength{\blanklineskip}
\newcommand{\hsindent}[1]{\quad}
\let\hspre\empty
\let\hspost\empty

\EndFmtInput
\makeatother
\ReadOnlyOnce{polycode.fmt}%
\makeatletter

\newcommand{\hsnewpar}[1]%
  {{\parskip=0pt\parindent=0pt\par\vskip #1\noindent}}

\newcommand{\hscodestyle}{}

\newcommand{\sethscode}[1]%
  {\expandafter\let\expandafter\hscode\csname #1\endcsname
   \expandafter\let\expandafter\endhscode\csname end#1\endcsname}

  {\par\noindent
   \advance\leftskip\mathindent
   \hscodestyle
   \let\\=\@normalcr
   \let\hspre\(\let\hspost\)%
   \pboxed}%
  {\endpboxed\)%
   \par\noindent
   \ignorespacesafterend}

  {\hsnewpar\abovedisplayskip
   \advance\leftskip\mathindent
   \hscodestyle
   \let\hspre\(\let\hspost\)%
   \pboxed}%
  {\endpboxed%
   \hsnewpar\belowdisplayskip
   \ignorespacesafterend}

  {\hsnewpar\abovedisplayskip
   \advance\leftskip\mathindent
   \hscodestyle
   \let\\=\@normalcr
   \(\pboxed}%
  {\endpboxed\)%
   \hsnewpar\belowdisplayskip
   \ignorespacesafterend}

\newcommand{\plainhs}{\sethscode{plainhscode}}

\plainhs

  {\hsnewpar\abovedisplayskip
   \advance\leftskip\mathindent
   \hscodestyle
   \let\\=\@normalcr
   \(\parray}%
  {\endparray\)%
   \hsnewpar\belowdisplayskip
   \ignorespacesafterend}

  {\parray}{\endparray}

  {\(\parray}{\endparray\)}

\def\codeframewidth{\arrayrulewidth}
\RequirePackage{calc}

  {\parskip=\abovedisplayskip\par\noindent
   \hscodestyle
   \arrayrulewidth=\codeframewidth
   \tabular{@{}|p{\linewidth-2\arraycolsep-2\arrayrulewidth-2pt}|@{}}%
   \hline\framedhslinecorrect\\{-1.5ex}%
   \let\endoflinesave=\\
   \let\\=\@normalcr
   \(\pboxed}%
  {\endpboxed\)%
   \framedhslinecorrect\endoflinesave{.5ex}\hline
   \endtabular
   \parskip=\belowdisplayskip\par\noindent
   \ignorespacesafterend}

\newcommand{\framedhslinecorrect}[2]%
  {#1[#2]}

  {\(\def\column##1##2{}%
   \let\>\undefined\let\<\undefined\let\\\undefined
   \newcommand\>[1][]{}\newcommand\<[1][]{}\newcommand\\[1][]{}%
   \def\fromto##1##2##3{##3}%
   }{\) }%

  {\let\orighscode=\hscode
   \let\origendhscode=\endhscode
   \def\endhscode{\def\hscode{\endgroup\def\@currenvir{hscode}\\}\begingroup}
   \orighscode\def\hscode{\endgroup\def\@currenvir{hscode}}}%
  {\origendhscode
   \global\let\hscode=\orighscode
   \global\let\endhscode=\origendhscode}%

\makeatother
\EndFmtInput

\RequirePackage[T1]{fontenc}
\RequirePackage[utf8x]{inputenc}
\RequirePackage{ucs}
\RequirePackage{amsfonts}

\DeclareUnicodeCharacter{737}{\textsuperscript{l}}
\DeclareUnicodeCharacter{8718}{\ensuremath{\blacksquare}}
\DeclareUnicodeCharacter{8759}{::}
\DeclareUnicodeCharacter{9669}{\ensuremath{\triangleleft}}
\DeclareUnicodeCharacter{8799}{\ensuremath{\stackrel{\scriptscriptstyle ?}{=}}}
\DeclareUnicodeCharacter{10214}{\ensuremath{\llbracket}}
\DeclareUnicodeCharacter{10215}{\ensuremath{\rrbracket}}

\let\Conid\Varid
\newcommand\Keyword[1]{\textsf{\textbf{#1}}}
\EndFmtInput

\begin{document}

\maketitle

\begin{abstract}

We present the approach underlying a course on
\emph{Domain-Specific Languages of Mathematics}
\cite{dslmcourseplan}, currently being developed at
Chalmers in response to difficulties faced by third-year students in
learning and applying classical mathematics (mainly real and complex
analysis).  The main idea is to encourage the students to approach
mathematical domains from a functional programming perspective: to
identify the main functions and types involved and, when necessary, to
introduce new abstractions; to give calculational proofs; to pay
attention to the syntax of the mathematical expressions; and, finally,
to organise the resulting functions and types in domain-specific
languages.

\end{abstract}

\section{Introduction}

In an article published in 2000 \cite{demoor2000pointwise}, de Moor
and Gibbons start by presenting an exam question for a first-year
course on algorithm design.  The question was not easy, but it also
did not seem particularly difficult.  Still:

\begin{quote}
  In the exam itself, however, no one got the answer right, so
  apparently this kind of question is too hard. That is discouraging,
  especially in view of the highly sophisticated problems that the
  same students can solve in mathematics exams.  Why is programming so
  much harder?
\end{quote}

Fifteen year later, we are confronted at Chalmers with the opposite
problem: many third-year students are having unusual difficulties in
courses involving classical mathematics (especially analysis, real and
complex) and its applications, while they seem quite capable of
dealing with ``highly sophisticated problems'' in computer science and
software engineering.  Why is mathematics so much harder?

One of the reasons for that is, we suspect, that by the third year
these students have grown very familiar with what could be called ``the
computer science perspective''.  For example, computer science places
strong emphasis on syntax and introduces conceptual tools for
describing it and resolving potential ambiguities.  In contrast to
this, mathematical notation is often ambiguous and context-dependent,
and there is no attempt to even make this ambiguity explicit (Sussman
and Wisdom talk about ``variables whose meaning depends upon and
changes with context, as well as the sort of impressionistic
mathematics that goes along with the use of such variables'', see
\cite{sussman2002role}).

Further, proofs in computer science tend to be more formal, often
using an equational logic format with explicit mention of the rules
that justify a given step, whereas mathematical proofs are presented
in natural language, with many steps being justified by an appeal to
intuition and to the semantical content, leaving a more precise
justification to the reader.  Unfortunately, the task of providing
such a justification requires a certain amount of expertise, and can
be discouraging to the beginner.

Mathematics requires (and rewards) active study.  Halmos, in a book
that cannot be strongly enough recommended, phrases it as follows
(\cite{halmos1985want}, page 69):

\begin{quote}
  It's been said before and often, but it cannot be overemphasised:
  study actively. Don't just read it; fight it!
\end{quote}

\noindent
but, as in the case of proofs, following this advice requires some
expertise, otherwise it risks being taken in too physical a sense.

In this paper, we present the approach underlying a course on
\emph{Domain-Specific Languages of Mathematics}
\cite{dslmcourseplan}, which is currently being developed at
Chalmers to alleviate these problems.  The main idea is to show the
students that they are, in fact, well-equipped to take an active
approach to mathematics: they need only apply the software engineering
and computer science tools they have acquired in the rest of their
studies.  The students should approach a mathematical domain in the
same way they would any other domain they are supposed to model as a
software system.

In particular, we are referring to the approach that a functional
programmer would take.  Functional programming deals with Modelling in
terms of types and pure functions, and this seems to be ideal for a
domain where functions are natural objects of study, and which is
possibly the only one where we can be certain that data is immutable.

Additionally, functional programming has, from the very beginning,
been connected to the notion of mathematical proof.  For example, the
influential language ML was originally developed in the 70s to be ``a
medium in which proofs \ldots can be expressed, as well as heuristic
algorithms for finding those proofs''
(\cite{biancuzzi2009masterminds}, page 205).

Explicitly introducing functions and their types, often left implicit
in mathematical texts, is an easy way to begin an active approach to
study.  Moreover, it serves as a way of relating new concepts to
familiar ones: even in continuous mathematics, many functions turn out
to be variants of the standard Haskell ones (not surprising,
considering that the former were often the inspiration for the
latter).  Finally, the explicit elements we introduce can be reasoned
about and lead to proofs in a more calculational style.  Section
\ref{sec:fandt} presents these elements in detail.

Section \ref{sec:dsls} deals with the higher-level question of the
organisation of our types and functions.  We emphasise
\emph{domain-specific languages} (DSLs, \cite{gibbons2013functional}),
since they are a good fit for the mathematical domain, which can
itself be seen as a collection of specialised languages.  Moreover,
building DSLs is increasingly becoming a standard industry practice
\cite{fowler2010domain}.  Empirical studies show that DSLs can lead to
fundamental increases in productivity, above alternative modelling
approaches such as UML \cite{tolvanen2011industrial}.  The course we
are developing will exercise and develop new skills in designing and
implementing DSLs.  The students will not simply use previously
acquired software engineering expertise, but also extend it, which can
be an important motivating aspect.

Both sections contain simple examples to illustrate our approach to an
active reading of mathematical texts.  The text we are reading is the
standard textbook used at Chalmers in the analysis course for first
year students (Adams and Essex, \cite{adams2010calculus}), though we
shall occasionally cite a few other texts as well.  At this stage, it
is important that we prevent a potentially grave misunderstanding of
our intentions.  We do not present the results of the active reading
as an ideal presentation of the mathematical concepts involved!  That
a presentation which is too explicit and complete can rob the readers
of a precious opportunity to exercise themselves is known to
mathematicians at least since Descartes' \emph{Geometry}
\cite{descartes1954geometry}:

\begin{quote}
  But I shall not stop to explain this in more detail, because I
  should deprive you of the pleasure of mastering it yourself, as well
  as of the advantage of training your mind by working over it, which
  is in my opinion the principal benefit to be derived from this
  science.
\end{quote}

On the other hand, the \emph{Geometry} was considered too obscure to
be read and didn't gain in popularity until van Schooten's explanatory
edition, so perhaps there is room for compromise.  In any case, both
mathematicians (\cite{wells1995communicating, kraft2004functions}) and
computer scientists (\cite{gries1995teaching, boute2009decibel}) have
argued that the computer science perspective could bring a valuable
contribution to mathematical education: we see our work as a step in
this direction.

We have been referring to the computer science students at Chalmers
since they are our main target audience, but we hope we can also
attract some of the mathematics students.  Indeed, for the latter the
course can serve as an introduction to functional programming and to
DSLs by means of examples with which they are familiar.  Thus,
ideally, the course would improve the mathematical education of
computer scientists and the computer science education of
mathematicians.

A word of warning.  We assume familiarity with Haskell (though not
with calculus), and we will take certain notational and semantic
liberties with it.  For example, we will use \ensuremath{\mathbin{:}} for the typing
relation, instead of \ensuremath{\Conid{::}}, and we will assume the existence of the
set-theoretical datatypes and operations used in classical analysis,
even though they are not implementable.  For example, we assume we
have at our disposal a powerset operation \ensuremath{\mathcal{P}}, (classical) real
numbers \ensuremath{\mathbb{R}}, choice operations, and so on.  We shall also use the
standard notation for intervals, which can lead to an overloading of
the Haskell list notation (\ensuremath{[\mskip1.5mu \Varid{a},\ \Varid{b}\mskip1.5mu]} may denote a closed interval or a
two-element list, depending on the context).

This paper, some associated source code and the DSLsofMath course
material is being collected on GitHub:
\url{https://github.com/DSLsofMath}. Contributions are welcome!

\section {Functions and types}
\label{sec:fandt}

One of the most useful actions of the student of a mathematical text
is to identify and type the functions involved.  If the notation she
uses is inadequate for this purpose, then her ability will be severely
impaired.  This is one of the main reasons for using functional
programming as the basis of our ``requirements engineering'' in a
mathematical domain.

Many important mathematical objects are functions.  Arguably, the
basic objects of study in undergraduate analysis are sequences of
one type or another.  Sequences are usually defined as functions of
positive integers (for example in Rudin \cite{rudin1976principles});
for the functional programmer it is perhaps more natural to model them
as functions of natural numbers, using \ensuremath{\Varid{a}\;\mathbin{:}\;\mathbb{N}\;\to \;\Conid{X}} where a
mathematician would write \ensuremath{\{\mskip1.5mu a_n\mskip1.5mu\}} or similar.
For brevity, we shall use \ensuremath{\Conid{X}} to denote a \ensuremath{\mathbb{R}} or \ensuremath{\mathbb{C}}, as is common
in undergraduate analysis, but in a classroom setting this could also
be an opportunity to explain type classes such as \ensuremath{\Conid{Num}}.

The notion of \emph{limit} is first defined for sequences.  The
operation of taking the limit is an example of a higher-order
function:
\begin{hscode}\SaveRestoreHook
\column{B}{@{}>{\hspre}l<{\hspost}@{}}%
\column{3}{@{}>{\hspre}l<{\hspost}@{}}%
\column{E}{@{}>{\hspre}l<{\hspost}@{}}%
\>[3]{}\Varid{lim}\;\mathbin{:}\;(\mathbb{N}\;\to \;\Conid{X})\;\to \;\Conid{X}{}\<[E]%
\ColumnHook
\end{hscode}\resethooks
Higher-order functions are ubiquitous in mathematical analysis, hence
the importance of using a notation that supports them in a simple way.
In fact, although we will not use it in this paper, it is often
necessary to account for \emph{dependent types}.  Intervals can, for
instance, be represented as dependent types, and all interval
operations are naturally dependently-typed.  In such cases, we would
prefer to use the notation of Agda \cite{norell2007towards,ionescu2013dependently} or Idris
\cite{brady2013idris}.

Convergent sequences can be used to represent real numbers, but the
use of sequences is much more diverse.  We can think of the sequence
of coefficients as a syntax that can be given multiple
interpretations:

\begin{itemize}
\item the sequence represents the coefficients of a series.  In this
  case, the semantics is usually given in terms of the limit (if it
  exists) of the sequence of partial sums:
\begin{hscode}\SaveRestoreHook
\column{B}{@{}>{\hspre}l<{\hspost}@{}}%
\column{4}{@{}>{\hspre}l<{\hspost}@{}}%
\column{13}{@{}>{\hspre}l<{\hspost}@{}}%
\column{16}{@{}>{\hspre}l<{\hspost}@{}}%
\column{23}{@{}>{\hspre}l<{\hspost}@{}}%
\column{30}{@{}>{\hspre}l<{\hspost}@{}}%
\column{E}{@{}>{\hspre}l<{\hspost}@{}}%
\>[4]{}\Sigma\;{}\<[13]%
\>[13]{}\mathbin{:}\;{}\<[16]%
\>[16]{}(\mathbb{N}\;\to \;\Conid{X})\;\to \;\Conid{X}{}\<[E]%
\\
\>[4]{}\Sigma\;\Varid{f}\;{}\<[13]%
\>[13]{}\mathrel{=}\;{}\<[16]%
\>[16]{}\Varid{lim}\;\Varid{s}\;{}\<[23]%
\>[23]{}\ \Keyword{where}\ \;{}\<[30]%
\>[30]{}\Varid{s}\;\Varid{n}\;\mathrel{=}\;\Varid{sum}\;(\Varid{map}\;\Varid{f}\;[\mskip1.5mu \Varid{0}\;\mathinner{\ldotp\ldotp}\;\Varid{n}\mskip1.5mu]){}\<[E]%
\ColumnHook
\end{hscode}\resethooks
\item the sequence represents the coefficients of a power series.  In
  this case, the semantics is that of a function, whose values are
  defined in terms of the evaluation of a series:
\begin{hscode}\SaveRestoreHook
\column{B}{@{}>{\hspre}l<{\hspost}@{}}%
\column{4}{@{}>{\hspre}l<{\hspost}@{}}%
\column{16}{@{}>{\hspre}l<{\hspost}@{}}%
\column{19}{@{}>{\hspre}l<{\hspost}@{}}%
\column{28}{@{}>{\hspre}l<{\hspost}@{}}%
\column{35}{@{}>{\hspre}l<{\hspost}@{}}%
\column{E}{@{}>{\hspre}l<{\hspost}@{}}%
\>[4]{}\Conid{Powers}\;{}\<[16]%
\>[16]{}\mathbin{:}\;{}\<[19]%
\>[19]{}(\mathbb{N}\;\to \;\Conid{X})\;\to \;(\Conid{X}\;\to \;\Conid{X}){}\<[E]%
\\
\>[4]{}\Conid{Powers}\;\Varid{a}\;\Varid{x}\;{}\<[16]%
\>[16]{}\mathrel{=}\;{}\<[19]%
\>[19]{}\Sigma\;\Varid{f}\;{}\<[28]%
\>[28]{}\ \Keyword{where}\ \;{}\<[35]%
\>[35]{}\Varid{f}\;\Varid{n}\;\mathrel{=}\;\Varid{a}\;\Varid{n}\;\Varid{*}\;\Varid{x}^{\Varid{n}}{}\<[E]%
\ColumnHook
\end{hscode}\resethooks
Power series are perhaps \emph{the} fundamental concept of
undergraduate analysis and its applications: they lead to elementary
and analytic functions, they are the starting point for the Fourier
and Laplace transformations, interval analysis, etc.  Therefore, the
student might find it puzzling that in most textbooks they do not have
a symbolic representation of their own, outside the somewhat unwieldy
$\sum\limits_{n=0}^\infty a_n X^n$.
\end{itemize}

The absence of explicit types in mathematical texts can sometimes lead
to confusing formulations.  For example, a standard text on
differential equations by Edwards, Penney and Calvis
\cite{edwards2008elementary} contains at page 266 the following
remark:

\begin{quote}
  The differentiation operator \ensuremath{\Conid{D}} can be viewed as a transformation
  which, when applied to the function \ensuremath{\Varid{f}\;(\Varid{t})}, yields the new function
  \ensuremath{\Conid{D}\;\{\mskip1.5mu \Varid{f}\;(\Varid{t})\mskip1.5mu\}\;\mathrel{=}\;\Varid{f'}\;(\Varid{t})}. The Laplace transformation \ensuremath{\mathcal{L}} involves the
  operation of integration and yields the new function \ensuremath{\mathcal{L}\;\{\mskip1.5mu \Varid{f}\;(\Varid{t})\mskip1.5mu\}\;\mathrel{=}\;\Conid{F}\;(\Varid{s})} of a new independent variable \ensuremath{\Varid{s}}.
\end{quote}

This is meant to introduce a distinction between ``operators'', such
as differentiation, which take functions to functions of the same
type, and ``transforms'', such as the Laplace transform, which take
functions to functions of a new type.  To the logician or the computer
scientist, the way of phrasing this difference in the quoted text
sounds strange: surely the \emph{name} of the independent variable
does not matter: the Laplace transformation could very well return a
function of the ``old'' variable \ensuremath{\Varid{t}}.  We can understand that the name
of the variable is used to carry semantic meaning about its type (this
is also common in functional programming, for example with the
conventional use of \ensuremath{\Varid{as}} to denote a list of \ensuremath{\Varid{a}}s).  Moreover, by
using this (implicit!) convention, it is easier to deal with cases
such as that of the Hartley transform, which does not change the type
of the input function, but rather the \emph{interpretation} of that
type.  We prefer to always give explicit typings rather than relying
on syntactical conventions, and to use type synonyms for the case in
which we have different interpretations of the same type.  In the
example of the Laplace transformation, this leads to
\begin{hscode}\SaveRestoreHook
\column{B}{@{}>{\hspre}l<{\hspost}@{}}%
\column{3}{@{}>{\hspre}l<{\hspost}@{}}%
\column{8}{@{}>{\hspre}l<{\hspost}@{}}%
\column{11}{@{}>{\hspre}l<{\hspost}@{}}%
\column{14}{@{}>{\hspre}l<{\hspost}@{}}%
\column{E}{@{}>{\hspre}l<{\hspost}@{}}%
\>[3]{}\Varid{type}\;\Conid{T}\;{}\<[11]%
\>[11]{}\mathrel{=}\;{}\<[14]%
\>[14]{}\mathbb{R}{}\<[E]%
\\
\>[3]{}\Varid{type}\;\Conid{S}\;{}\<[11]%
\>[11]{}\mathrel{=}\;{}\<[14]%
\>[14]{}\mathbb{C}{}\<[E]%
\\
\>[3]{}\mathcal{L}\;{}\<[8]%
\>[8]{}\mathbin{:}\;{}\<[11]%
\>[11]{}(\Conid{T}\;\to \;\mathbb{C})\;\to \;(\Conid{S}\;\to \;\mathbb{C}){}\<[E]%
\ColumnHook
\end{hscode}\resethooks
In the following subsection, we present two simple examples of
``close reading'' a mathematical text, trying to identify and type the
functions involved, and to relate them to the familiar elements of
functional programming.

\subsection{Two examples}
\label{subsec:twoexamples}

Consider the following statement of the completeness property for
\ensuremath{\mathbb{R}} (\cite{adams2010calculus}, page 4):

\begin{quote}

    The \emph{completeness} property of the real number system is more
    subtle and difficult to understand. One way to state it is as
    follows: if \ensuremath{\Conid{A}} is any set of real numbers having at least one
    number in it, and if there exists a real number \ensuremath{\Varid{y}} with the
    property that \ensuremath{\Varid{x}\;\leq \;\Varid{y}} for every \ensuremath{\Varid{x}\;\in\;\Conid{A}} (such a number \ensuremath{\Varid{y}} is
    called an \textbf{upper bound} for \ensuremath{\Conid{A}}), then there exists a
    smallest such number, called the \textbf{least upper bound} or
    \textbf{supremum} of \ensuremath{\Conid{A}}, and denoted \ensuremath{\Varid{sup}\;(\Conid{A})}. Roughly speaking,
    this says that there can be no holes or gaps on the real
    line---every point corresponds to a real number.

\end{quote}

The functional programmer trying to make sense of this ``subtle and
difficult to understand'' property will start by making explicit the
functions involved:
\begin{hscode}\SaveRestoreHook
\column{B}{@{}>{\hspre}l<{\hspost}@{}}%
\column{3}{@{}>{\hspre}l<{\hspost}@{}}%
\column{E}{@{}>{\hspre}l<{\hspost}@{}}%
\>[3]{}\Varid{sup}\;\mathbin{:}\;\mathcal{P}^+\;\mathbb{R}\;\to \;\mathbb{R}{}\<[E]%
\ColumnHook
\end{hscode}\resethooks
\ensuremath{\Varid{sup}} is defined only for those subsets of \ensuremath{\mathbb{R}} which are bounded
from above; for these it returns the least upper bound.

Functional programmers are acquainted with a large number of standard
functions.  Among these are \ensuremath{\Varid{minimum}} and \ensuremath{\Varid{maximum}}, which
return the smallest and the largest element of a given (non-empty)
list.  It is easy enough to specify set versions of these functions,
for example:
\begin{hscode}\SaveRestoreHook
\column{B}{@{}>{\hspre}l<{\hspost}@{}}%
\column{3}{@{}>{\hspre}l<{\hspost}@{}}%
\column{10}{@{}>{\hspre}l<{\hspost}@{}}%
\column{13}{@{}>{\hspre}l<{\hspost}@{}}%
\column{16}{@{}>{\hspre}l<{\hspost}@{}}%
\column{29}{@{}>{\hspre}l<{\hspost}@{}}%
\column{E}{@{}>{\hspre}l<{\hspost}@{}}%
\>[3]{}\Varid{min}\;{}\<[10]%
\>[10]{}\mathbin{:}\;{}\<[13]%
\>[13]{}\mathcal{P}^+\;\mathbb{R}\;\to \;\mathbb{R}{}\<[E]%
\\
\>[3]{}\Varid{min}\;\Conid{A}\;{}\<[10]%
\>[10]{}\mathrel{=}\;{}\<[13]%
\>[13]{}\Varid{x}\;{}\<[16]%
\>[16]{}\iff\;{}\<[29]%
\>[29]{}(\Varid{x}\;\in\;\Conid{A})\;\mathrel{\wedge}\;(\forall\ \Varid{a}\;\in\;\Conid{A}.\ \ \Varid{x}\;\leq \;\Varid{a}){}\<[E]%
\ColumnHook
\end{hscode}\resethooks
\ensuremath{\Varid{min}} on sets enjoys similar properties to its list counterpart, and
some are easier to prove in this context, since the structure is
simpler (no duplicates, no ordering of elements).  For
example, we have

\begin{quote}
  If \ensuremath{\Varid{y}\;\Varid{<}\;\Varid{min}\;\Conid{A}}, then \ensuremath{\Varid{y}\;\notin\;\Conid{A}}.
\end{quote}

Exploring the relationship between the ``new'' function \ensuremath{\Varid{sup}} and the
familiar \ensuremath{\Varid{min}} and \ensuremath{\Varid{max}} can dispel some of the difficulties involved
in the completeness property.  For example, \ensuremath{\Varid{sup}\;\Conid{A}} is similar to \ensuremath{\Varid{max}\;\Conid{A}}: if the latter is defined, then so is the former, and they are
equal.  But \ensuremath{\Varid{sup}\;\Conid{A}} is also the smallest element of a set, which
suggests a connection to \ensuremath{\Varid{min}}.  To see this, introduce the function
\begin{hscode}\SaveRestoreHook
\column{B}{@{}>{\hspre}l<{\hspost}@{}}%
\column{3}{@{}>{\hspre}l<{\hspost}@{}}%
\column{10}{@{}>{\hspre}l<{\hspost}@{}}%
\column{13}{@{}>{\hspre}l<{\hspost}@{}}%
\column{E}{@{}>{\hspre}l<{\hspost}@{}}%
\>[3]{}\Varid{ubs}\;{}\<[10]%
\>[10]{}\mathbin{:}\;{}\<[13]%
\>[13]{}\mathcal{P}\;\mathbb{R}\;\to \;\mathcal{P}\;\mathbb{R}{}\<[E]%
\\
\>[3]{}\Varid{ubs}\;\Conid{A}\;{}\<[10]%
\>[10]{}\mathrel{=}\;{}\<[13]%
\>[13]{}\{\mskip1.5mu \Varid{x}\;\mid \;\Varid{x}\;\in\;\mathbb{R},\ \Varid{x}\;\Varid{upper}\;\Varid{bound}\;\Varid{of}\;\Conid{A}\mskip1.5mu\}\;{}\<[E]%
\\
\>[10]{}\mathrel{=}\;{}\<[13]%
\>[13]{}\{\mskip1.5mu \Varid{x}\;\mid \;\Varid{x}\;\in\;\mathbb{R},\ \forall\ \Varid{a}\;\in\;\Conid{A}.\ \ \Varid{a}\;\leq \;\Varid{x}\mskip1.5mu\}{}\<[E]%
\ColumnHook
\end{hscode}\resethooks
which returns the set of upper bounds of \ensuremath{\Conid{A}}.  The completeness axiom
can be stated as

\begin{quote}
  Assume an \ensuremath{\Conid{A}\;\mathbin{:}\;\mathcal{P}^+\;\mathbb{R}} with an upper bound \ensuremath{\Varid{u}\;\in\;\Varid{ubs}\;\Conid{A}}.

  Then \ensuremath{\Varid{s}\;\mathrel{=}\;\Varid{sup}\;\Conid{A}\;\mathrel{=}\;\Varid{min}\;(\Varid{ubs}\;\Conid{A})} exists.
\end{quote}

\noindent
where
\begin{hscode}\SaveRestoreHook
\column{B}{@{}>{\hspre}l<{\hspost}@{}}%
\column{3}{@{}>{\hspre}l<{\hspost}@{}}%
\column{E}{@{}>{\hspre}l<{\hspost}@{}}%
\>[3]{}\Varid{sup}\;\mathbin{:}\;\mathcal{P}^+\;\mathbb{R}\;\to \;\mathbb{R}{}\<[E]%
\\
\>[3]{}\Varid{sup}\;\mathrel{=}\;\Varid{min}\;\mathbin{\circ}\;\Varid{ubs}{}\<[E]%
\ColumnHook
\end{hscode}\resethooks
So, now we know that for any bounded set \ensuremath{\Conid{A}} we have a supremum \ensuremath{\Varid{s}\;\mathbin{:}\;\mathbb{R}}, but \ensuremath{\Varid{s}} need not be in \ensuremath{\Conid{A}} --- could there be a ``gap''?
(An example set could be \ensuremath{\Conid{A}\;\mathrel{=}\;\{\mskip1.5mu \Varid{7}\;\Varid{-}\;1/n\;\mid \;\Varid{n}\;\in\;\mathbb{N}^+\mskip1.5mu\}} with \ensuremath{\Varid{s}\;\mathrel{=}\;\Varid{sup}\;\Conid{A}\;\mathrel{=}\;\Varid{7}\;\notin\;\Conid{A}}.)
If we by ``gap'' mean ``an \ensuremath{\epsilon}-neighbourhood between \ensuremath{\Conid{A}} and \ensuremath{\Varid{s}}''
we can prove there is in fact no ``gap''.

The explicit introduction of functions such as \ensuremath{\Varid{ubs}} allows us to give
calculational proofs in the style introduced by Wim Feijen and used in
many computer science textbooks, especially in functional programming
(such proofs are more amenable to automatic verification, see for
example the algebra of programming library implemented in Agda
\cite{mu2009algebra}).  For example, if \ensuremath{\Varid{s}\;\mathrel{=}\;\Varid{sup}\;\Conid{A}}:

\def\commentbegin{\quad\{\ }
\def\commentend{\}}

\begin{hscode}\SaveRestoreHook
\column{B}{@{}>{\hspre}l<{\hspost}@{}}%
\column{4}{@{}>{\hspre}l<{\hspost}@{}}%
\column{E}{@{}>{\hspre}l<{\hspost}@{}}%
\>[4]{}\Varid{0}\;\Varid{<}\;\epsilon{}\<[E]%
\\[\blanklineskip]%
\>[B]{}\Rightarrow \;\mbox{\commentbegin  arithmetic  \commentend}{}\<[E]%
\\[\blanklineskip]%
\>[B]{}\hsindent{4}{}\<[4]%
\>[4]{}\Varid{s}\;\Varid{-}\;\epsilon\;\Varid{<}\;\Varid{s}{}\<[E]%
\\[\blanklineskip]%
\>[B]{}\Rightarrow \;\mbox{\commentbegin  \ensuremath{\Varid{s}\;\mathrel{=}\;\Varid{min}\;(\Varid{ubs}\;\Conid{A})}, property of \ensuremath{\Varid{min}} from above  \commentend}{}\<[E]%
\\[\blanklineskip]%
\>[B]{}\hsindent{4}{}\<[4]%
\>[4]{}\Varid{s}\;\Varid{-}\;\epsilon\;\notin\;\Varid{ubs}\;\Conid{A}{}\<[E]%
\\[\blanklineskip]%
\>[B]{}\Rightarrow \;\mbox{\commentbegin  set membership  \commentend}{}\<[E]%
\\[\blanklineskip]%
\>[B]{}\hsindent{4}{}\<[4]%
\>[4]{}\neg \;\forall\ \Varid{a}\;\in\;\Conid{A}.\ \ \Varid{a}\;\leq \;\Varid{s}\;\Varid{-}\;\epsilon{}\<[E]%
\\[\blanklineskip]%
\>[B]{}\Rightarrow \;\mbox{\commentbegin  quantifier negation  \commentend}{}\<[E]%
\\[\blanklineskip]%
\>[B]{}\hsindent{4}{}\<[4]%
\>[4]{}\exists\ \Varid{a}\;\in\;\Conid{A}.\ \ \Varid{s}\;\Varid{-}\;\epsilon\;\Varid{<}\;\Varid{a}{}\<[E]%
\\[\blanklineskip]%
\>[B]{}\Rightarrow \;\mbox{\commentbegin  definition of upper bound  \commentend}{}\<[E]%
\\[\blanklineskip]%
\>[B]{}\hsindent{4}{}\<[4]%
\>[4]{}\exists\ \Varid{a}\;\in\;\Conid{A}.\ \ \Varid{s}\;\Varid{-}\;\epsilon\;\Varid{<}\;\Varid{a}\;\leq \;\Varid{s}{}\<[E]%
\\[\blanklineskip]%
\>[B]{}\Rightarrow \;\mbox{\commentbegin  subtract \ensuremath{\Varid{s}}, use \ensuremath{\Varid{0}\;\Varid{<}\;\epsilon}  \commentend}{}\<[E]%
\\[\blanklineskip]%
\>[B]{}\hsindent{4}{}\<[4]%
\>[4]{}\exists\ \Varid{a}\;\in\;\Conid{A}.\ \ \Varid{-}\;\epsilon\;\Varid{<}\;\Varid{a}\;\Varid{-}\;\Varid{s}\;\Varid{<}\;\epsilon{}\<[E]%
\\[\blanklineskip]%
\>[B]{}\Rightarrow \;\mbox{\commentbegin  absolute value  \commentend}{}\<[E]%
\\[\blanklineskip]%
\>[B]{}\hsindent{4}{}\<[4]%
\>[4]{}\exists\ \Varid{a}\;\in\;\Conid{A}.\ \ (\lvert{}\Varid{a}\;\Varid{-}\;\Varid{s}\rvert{}\;\Varid{<}\;\epsilon){}\<[E]%
\\[\blanklineskip]%
\>[B]{}\Rightarrow \;\mbox{\commentbegin  introduce the neighbourhood function \ensuremath{\Conid{V}\;\mathbin{:}\;\Conid{X}\;\to \;\mathbb{R}_{> 0}\;\to \;\mathcal{P}\;\Conid{X}}  \commentend}{}\<[E]%
\\[\blanklineskip]%
\>[B]{}\hsindent{4}{}\<[4]%
\>[4]{}\exists\ \Varid{a}\;\in\;\Conid{A}.\ \ \Varid{a}\;\in\;\Conid{V}\;\Varid{s}\;\epsilon{}\<[E]%
\ColumnHook
\end{hscode}\resethooks

This simple proof shows that we can always find an element of \ensuremath{\Conid{A}} as
near to \ensuremath{\Varid{sup}\;\Conid{A}} as we want, which explains perhaps the above statement
``Roughly speaking, [the completeness axiom] says that there can be no
holes or gaps on the real line---every point corresponds to a real
number.''


As another example of work on the text, consider the following
definition (\cite{adams2010calculus}, page A-23):

\begin{quote}
  \textbf{Limit of a sequence}

  We say that \ensuremath{\Varid{lim}\;x_n\;\mathrel{=}\;\Conid{L}} if for every positive number \ensuremath{\epsilon}
  there exists a positive number \ensuremath{\Conid{N}\;\mathrel{=}\;\Conid{N}\;(\epsilon)} such that
  \ensuremath{\lvert{}x_n\;\Varid{-}\;\Conid{L}\rvert{}\;\Varid{<}\;\epsilon} holds whenever \ensuremath{\Varid{n}\;\geq \;\Conid{N}}.
\end{quote}

\noindent
There are many opportunities for functional programmers to apply their
craft here, such as

\begin{itemize}
\item giving an explicit typing \ensuremath{\Varid{lim}\;\mathbin{:}\;(\mathbb{N}\;\to \;\Conid{X})\;\to \;\Conid{X}} and writing
  \ensuremath{\Varid{lim}\;\Varid{x}} in order to avoid the impression that the result depends on
  some particular value \ensuremath{x_n};
\item giving an explicit typing for the absolute value function \ensuremath{\lvert{}\anonymous \rvert{}\;\mathbin{:}\;\Conid{X}\;\to \;\mathbb{R}_{\ge 0}};
\item introducing explicitly the function \ensuremath{\Conid{N}\;\mathbin{:}\;\mathbb{R}_{> 0}\;\to \;\mathbb{N}};
\item introducing a neighbourhood function \ensuremath{\Conid{V}\;\mathbin{:}\;\Conid{X}\;\to \;\mathbb{R}_{> 0}\;\to \;\mathcal{P}\;\Conid{X}} with
\begin{hscode}\SaveRestoreHook
\column{B}{@{}>{\hspre}l<{\hspost}@{}}%
\column{4}{@{}>{\hspre}l<{\hspost}@{}}%
\column{E}{@{}>{\hspre}l<{\hspost}@{}}%
\>[4]{}\Conid{V}\;\Varid{x}\;\epsilon\;\mathrel{=}\;\{\mskip1.5mu \Varid{x'}\;\mid \;\Varid{x'}\;\in\;\Conid{X},\ \lvert{}\Varid{x'}\;\Varid{-}\;\Varid{x}\rvert{}\;\Varid{<}\;\epsilon\mskip1.5mu\}{}\<[E]%
\ColumnHook
\end{hscode}\resethooks
\end{itemize}

These are all just changes in the notation of elements already present
in the text (the \emph{neighbourhood} function \ensuremath{\Conid{V}} is introduced in
Adams, but first on page 567, long after the chapter on sequences and
convergence, page 495).  Many real analysis textbooks adopt, in fact,
the one or the other of these changes.  However, functional
programmers will probably observe that the expression \ensuremath{a_n} \ldots\
\emph{whenever} \ensuremath{\Varid{n}\;\geq \;\Conid{N}} refers to the \ensuremath{\Conid{N}}th tail of the sequence,
i.e., to the elements remaining after the first \ensuremath{\Conid{N}} elements have been
dropped.  This recalls the familiar Haskell function \ensuremath{\Varid{drop}\;\mathbin{:}\;\Conid{Int}\;\to \;[\mskip1.5mu \Varid{a}\mskip1.5mu]\;\to \;[\mskip1.5mu \Varid{a}\mskip1.5mu]}, which can be recast to suit the new context:
\begin{hscode}\SaveRestoreHook
\column{B}{@{}>{\hspre}l<{\hspost}@{}}%
\column{3}{@{}>{\hspre}l<{\hspost}@{}}%
\column{13}{@{}>{\hspre}l<{\hspost}@{}}%
\column{16}{@{}>{\hspre}l<{\hspost}@{}}%
\column{29}{@{}>{\hspre}l<{\hspost}@{}}%
\column{E}{@{}>{\hspre}l<{\hspost}@{}}%
\>[3]{}\Varid{drop}\;\mathbin{:}\;\mathbb{N}\;\to \;(\mathbb{N}\;\to \;\Conid{X})\;\to \;(\mathbb{N}\;\to \;\Conid{X}){}\<[E]%
\\
\>[3]{}\Varid{drop}\;\Varid{n}\;\Varid{f}\;\mathrel{=}\;\lambda \;(\Varid{i}\;\mathbin{:}\;\mathbb{N})\;\to \;{}\<[29]%
\>[29]{}\Varid{f}\;(\Varid{n}\;+\;\Varid{i}){}\<[E]%
\\[\blanklineskip]%
\>[3]{}\Conid{Drop}\;\mathbin{:}\;\mathbb{N}\;\to \;(\mathbb{N}\;\to \;\Conid{X})\;\to \;\mathcal{P}\;\Conid{X}{}\<[E]%
\\
\>[3]{}\Conid{Drop}\;\Varid{n}\;\Varid{f}\;{}\<[13]%
\>[13]{}\mathrel{=}\;{}\<[16]%
\>[16]{}\Varid{range}\;(\Varid{drop}\;\Varid{n}\;\Varid{f})\;{}\<[E]%
\\
\>[13]{}\mathrel{=}\;{}\<[16]%
\>[16]{}\{\mskip1.5mu \Varid{f}\;\Varid{i}\;\mid \;\Varid{i}\;\in\;\mathbb{N},\ \Varid{n}\;\leq \;\Varid{i}\mskip1.5mu\}{}\<[E]%
\ColumnHook
\end{hscode}\resethooks
The function \ensuremath{\Conid{Drop}} has many properties, for example:

\begin{itemize}
\item anti-monotone in the first argument
\begin{hscode}\SaveRestoreHook
\column{B}{@{}>{\hspre}l<{\hspost}@{}}%
\column{3}{@{}>{\hspre}l<{\hspost}@{}}%
\column{E}{@{}>{\hspre}l<{\hspost}@{}}%
\>[3]{}\Varid{m}\;\leq \;\Varid{n}\;\Rightarrow \;\Conid{Drop}\;\Varid{n}\;\Varid{f}\;\subseteq\;\Conid{Drop}\;\Varid{m}\;\Varid{f}{}\<[E]%
\ColumnHook
\end{hscode}\resethooks
in particular \ensuremath{\Conid{Drop}\;\Varid{n}\;\Varid{f}\;\subseteq\;\Conid{Drop}\;\Varid{0}\;\Varid{f}} for
  all \ensuremath{\Varid{n}};

\item if \ensuremath{\Varid{f}} is increasing, then, for any \ensuremath{\Varid{m}} and \ensuremath{\Varid{n}}
\begin{hscode}\SaveRestoreHook
\column{B}{@{}>{\hspre}l<{\hspost}@{}}%
\column{3}{@{}>{\hspre}l<{\hspost}@{}}%
\column{E}{@{}>{\hspre}l<{\hspost}@{}}%
\>[3]{}\Varid{ubs}\;(\Conid{Drop}\;\Varid{m}\;\Varid{f})\;\mathrel{=}\;\Varid{ubs}\;(\Conid{Drop}\;\Varid{n}\;\Varid{f}){}\<[E]%
\ColumnHook
\end{hscode}\resethooks
and therefore, if \ensuremath{\Conid{Drop}\;\Varid{0}\;\Varid{f}} is bounded
\begin{hscode}\SaveRestoreHook
\column{B}{@{}>{\hspre}l<{\hspost}@{}}%
\column{3}{@{}>{\hspre}l<{\hspost}@{}}%
\column{E}{@{}>{\hspre}l<{\hspost}@{}}%
\>[3]{}\Varid{sup}\;(\Conid{Drop}\;\Varid{m}\;\Varid{f})\;\mathrel{=}\;\Varid{sup}\;(\Conid{Drop}\;\Varid{n}\;\Varid{f}){}\<[E]%
\ColumnHook
\end{hscode}\resethooks
\item if \ensuremath{\Varid{f}} is increasing, then
\begin{hscode}\SaveRestoreHook
\column{B}{@{}>{\hspre}l<{\hspost}@{}}%
\column{3}{@{}>{\hspre}l<{\hspost}@{}}%
\column{E}{@{}>{\hspre}l<{\hspost}@{}}%
\>[3]{}\Conid{Drop}\;\Varid{n}\;\Varid{f}\;\subseteq\;[\Varid{f}\;\Varid{n},\ \infty){}\<[E]%
\ColumnHook
\end{hscode}\resethooks
\end{itemize}

\noindent
Using \ensuremath{\Conid{Drop}}, we have that

\begin{hscode}\SaveRestoreHook
\column{B}{@{}>{\hspre}l<{\hspost}@{}}%
\column{3}{@{}>{\hspre}l<{\hspost}@{}}%
\column{30}{@{}>{\hspre}l<{\hspost}@{}}%
\column{E}{@{}>{\hspre}l<{\hspost}@{}}%
\>[3]{}\Varid{lim}\;\Varid{f}\;\mathrel{=}\;\Conid{L}{}\<[E]%
\\
\>[B]{}\iff\;{}\<[E]%
\\
\>[B]{}\hsindent{3}{}\<[3]%
\>[3]{}\exists\ \Conid{N}\;\mathbin{:}\;\mathbb{R}_{> 0}\;\to \;\mathbb{N}.\ \ {}\<[30]%
\>[30]{}\forall\ \epsilon\;\in\;\mathbb{R}_{> 0}.\ \ \Conid{Drop}\;(\Conid{N}\;\epsilon)\;\Varid{f}\;\subseteq\;\Conid{V}\;\Conid{L}\;\epsilon{}\<[E]%
\ColumnHook
\end{hscode}\resethooks

This formulation has the advantage of eliminating one of the three
quantifiers in the definition of limit.  In general, introducing
functions and operations on functions leads to fewer quantifiers.  For
example, we could lift inclusion of sets to the function level: for
\ensuremath{\Varid{f},\ \Varid{g}\;\mathbin{:}\;\Conid{A}\;\to \;\mathcal{P}\;\Conid{B}} define
\begin{hscode}\SaveRestoreHook
\column{B}{@{}>{\hspre}l<{\hspost}@{}}%
\column{3}{@{}>{\hspre}l<{\hspost}@{}}%
\column{19}{@{}>{\hspre}l<{\hspost}@{}}%
\column{32}{@{}>{\hspre}l<{\hspost}@{}}%
\column{54}{@{}>{\hspre}l<{\hspost}@{}}%
\column{60}{@{}>{\hspre}l<{\hspost}@{}}%
\column{70}{@{}>{\hspre}l<{\hspost}@{}}%
\column{E}{@{}>{\hspre}l<{\hspost}@{}}%
\>[3]{}\Varid{f}\;\subseteq\;\Varid{g}\;{}\<[19]%
\>[19]{}\iff\;{}\<[32]%
\>[32]{}\forall\ \Varid{a}\;\in\;\Conid{A}.\ \ {}\<[54]%
\>[54]{}\Varid{f}\;\Varid{a}\;{}\<[60]%
\>[60]{}\subseteq\;{}\<[70]%
\>[70]{}\Varid{g}\;\Varid{a}{}\<[E]%
\ColumnHook
\end{hscode}\resethooks
and we could eliminate the quantification of \ensuremath{\epsilon} above:

\begin{hscode}\SaveRestoreHook
\column{B}{@{}>{\hspre}l<{\hspost}@{}}%
\column{5}{@{}>{\hspre}l<{\hspost}@{}}%
\column{32}{@{}>{\hspre}l<{\hspost}@{}}%
\column{E}{@{}>{\hspre}l<{\hspost}@{}}%
\>[5]{}\exists\ \Conid{N}\;\mathbin{:}\;\mathbb{R}_{> 0}\;\to \;\mathbb{N}.\ \ {}\<[32]%
\>[32]{}\forall\ \epsilon\;\in\;\mathbb{R}_{> 0}.\ \ \Conid{Drop}\;(\Conid{N}\;\epsilon)\;\Varid{f}\;\subseteq\;\Conid{V}\;\Conid{L}\;\epsilon{}\<[E]%
\\
\>[B]{}\iff\;{}\<[E]%
\\
\>[B]{}\hsindent{5}{}\<[5]%
\>[5]{}\exists\ \Conid{N}\;\mathbin{:}\;\mathbb{R}_{> 0}\;\to \;\mathbb{N}.\ \ {}\<[32]%
\>[32]{}(\Varid{flip}\;\Conid{Drop}\;\Varid{f}\;\mathbin{\circ}\;\Conid{N})\;\subseteq\;\Conid{V}\;\Conid{L}{}\<[E]%
\ColumnHook
\end{hscode}\resethooks

The application of \ensuremath{\Varid{flip}} is necessary to bring the arguments in the
correct order.  As this example shows, sometimes the price of
eliminating quantifiers can be too high.

We can show that increasing sequences which are bounded from above are
convergent.
Let \ensuremath{\Varid{f}} be a sequence bounded from above (i.e., with \ensuremath{\Conid{A}\;\mathrel{=}\;\Conid{Drop}\;\Varid{0}\;\Varid{f}}
there is some \ensuremath{\Varid{u}\;\in\;\Varid{ubs}\;\Conid{A}}), and let \ensuremath{\Varid{s}\;\mathrel{=}\;\Varid{sup}\;\Conid{A}}.
Then, we know from our previous example that \ensuremath{\exists\ \Varid{a}\;\in\;\Conid{A}.\ \ \Varid{a}\;\in\;\Conid{V}\;\Varid{s}\;\epsilon} for any \ensuremath{\epsilon}.
Or equivalently, \ensuremath{\forall\ \epsilon\;\in\;\mathbb{R}_{> 0}.\ \ \exists\ \Varid{i}\;\in\;\mathbb{N}.\ \ \Varid{f}\;\Varid{i}\;\in\;\Conid{V}\;\Varid{s}\;\epsilon}.
Finally by swapping quantifier order and introducing the name \ensuremath{\Conid{N}} for
the function that determines \ensuremath{\Varid{i}} from \ensuremath{\epsilon} we obtain \ensuremath{\exists\ \Conid{N}\;\mathbin{:}\;\mathbb{R}_{> 0}\;\to \;\mathbb{N}.\ \ \Varid{f}\;(\Conid{N}\;\epsilon)\;\in\;\Conid{V}\;\Varid{s}\;\epsilon}.

If \ensuremath{\Varid{f}} is increasing, we have

\begin{hscode}\SaveRestoreHook
\column{B}{@{}>{\hspre}l<{\hspost}@{}}%
\column{3}{@{}>{\hspre}l<{\hspost}@{}}%
\column{E}{@{}>{\hspre}l<{\hspost}@{}}%
\>[3]{}\Conid{Drop}\;(\Conid{N}\;\epsilon)\;\Varid{f}{}\<[E]%
\\
\>[B]{}\subseteq\;\mbox{\commentbegin  \ensuremath{\Varid{f}} increasing  \commentend}{}\<[E]%
\\
\>[B]{}\hsindent{3}{}\<[3]%
\>[3]{}[\mskip1.5mu \Varid{f}\;(\Conid{N}\;\epsilon),\ \Varid{sup}\;(\Conid{Drop}\;(\Conid{N}\;\epsilon)\;\Varid{f})\mskip1.5mu]{}\<[E]%
\\
\>[B]{}\mathrel{=}\;\mbox{\commentbegin  \ensuremath{\Varid{f}} increasing \ensuremath{\Rightarrow \;\Varid{sup}\;(\Conid{Drop}\;\Varid{n}\;\Varid{f})\;\mathrel{=}\;\Varid{sup}\;(\Conid{Drop}\;\Varid{0}\;\Varid{f})\;\mathrel{=}\;\Varid{s}}  \commentend}{}\<[E]%
\\
\>[B]{}\hsindent{3}{}\<[3]%
\>[3]{}[\mskip1.5mu \Varid{f}\;(\Conid{N}\;\epsilon),\ \Varid{s}\mskip1.5mu]{}\<[E]%
\\
\>[B]{}\subseteq\;\mbox{\commentbegin  \ensuremath{\Varid{f}\;(\Conid{N}\;\epsilon)\;\in\;\Conid{V}\;\Varid{s}\;\epsilon}  \commentend}{}\<[E]%
\\
\>[B]{}\hsindent{3}{}\<[3]%
\>[3]{}\Conid{V}\;\Varid{s}\;\epsilon{}\<[E]%
\ColumnHook
\end{hscode}\resethooks

As before, the introduction of a new function has helped in relating
familiar elements (the standard Haskell function \ensuremath{\Varid{drop}}) to new ones
(the concept of limit) and to formulate proofs in a calculational
style.

\section{Domain-specific languages}
\label{sec:dsls}

There is no clear-cut line between libraries and DSLs, and intuitions
differ.  For example, in Chapter 8 of \emph{Thinking Functionally with
  Haskell} (\cite{bird2014thinking}), Richard Bird presents a language
for pretty-printing documents based on Wadler's chapter in \emph{The
  Fun of Programming} \cite{wadler2003prettier}, but refers to it as a
library, only mentioning DSLs in the chapter notes.

Both libraries and DSLs are collections of types and functions meant
to represent concepts from a domain at a high level of abstraction.
What separates a DSL from a library is, in our opinion, the deliberate
separation of syntax from semantics, which is a feature of all
programming languages (and, arguably, of languages in general).

As we have seen above, in mathematics the syntactical elements are
sometimes conflated with the semantical ones ($f(t)$ versus $f(s)$,
for example), and disentangling the two aspects can be an important
aid in coming to terms with a mathematical text.  Hence, our emphasis
on DSLs rather than libraries.

The distinction between syntax and semantics is, in fact, quite common
in mathematics, often hiding behind the keyword ``formal''.  For
example, \emph{formal power series} are an attempt to present the
theory of power series restricted to their syntactic aspects,
independent of their semantic interpretations in terms of convergence
(in the various domains of real numbers, complex numbers, intervals of
reals, etc.).  The ``formalist'' texts of Bourbaki present various
domains of mathematics by emphasising their formal properties
(\emph{axiomatic structure}), then relating those in terms of ``lower
levels'', with the lowest levels expressed in terms of set theory (so,
for example, groups are initially introduced axiomatically, then
various interpretations are discussed, such as ``groups of
transformations'', which in turn are interpreted in terms of
endo-functions, which are ultimately represented as sets of ordered
pairs).  Currently, however, even the most ``formalist'' mathematical
texts offer to the computer scientist many opportunities for active
reading.

\subsection{A case study: complex numbers}

To illustrate the above, we present an analytic reading of the
introduction of complex numbers in \cite{adams2010calculus}.  The
simplicity of the domain is meant to allow the reader to concentrate
on the essential elements of our approach without the distraction of
potentially unfamiliar mathematical concepts.  Because of the
exemplary character of this section, we bracket our previous knowledge
and approach the text as we would a completely new domain, even if
that leads to a somewhat exaggerated attention to detail.

Adams and Essex introduce complex numbers in Appendix 1.  The
section \emph{Definition of Complex Numbers} begins with:

\begin{quote}
  We begin by defining the symbol \ensuremath{\Varid{i}}, called \textbf{the imaginary unit}, to
  have the property
\begin{hscode}\SaveRestoreHook
\column{B}{@{}>{\hspre}l<{\hspost}@{}}%
\column{8}{@{}>{\hspre}l<{\hspost}@{}}%
\column{E}{@{}>{\hspre}l<{\hspost}@{}}%
\>[8]{}\Varid{i}^2\;\mathrel{=}\;\Varid{-1}{}\<[E]%
\ColumnHook
\end{hscode}\resethooks
  Thus, we could also call \ensuremath{\Varid{i}} the square root of \ensuremath{\Varid{-1}} and denote it
  \ensuremath{\sqrt{\Varid{-1}}}. Of course, \ensuremath{\Varid{i}} is not a real number; no real number has
  a negative square.
\end{quote}

At this stage, it is not clear what the type of \ensuremath{\Varid{i}} is meant to be, we
only know that \ensuremath{\Varid{i}} is not a real number.  Moreover, we do not know
what operations are possible on \ensuremath{\Varid{i}}, only that \ensuremath{\Varid{i}^2} is another name
for \ensuremath{\Varid{-1}} (but it is not obvious that, say \ensuremath{\Varid{i}\;\Varid{*}\;\Varid{i}} is related in
any way with \ensuremath{\Varid{i}^2}, since the operations of multiplication and
squaring have only been introduced so far for numerical types such as
\ensuremath{\mathbb{N}} or \ensuremath{\mathbb{R}}, and not for symbols).

For the moment, we introduce a type for the value \ensuremath{\Varid{i}}, and, since we
know nothing about other values, we make \ensuremath{\Varid{i}} the only member of this
type:
\begin{hscode}\SaveRestoreHook
\column{B}{@{}>{\hspre}l<{\hspost}@{}}%
\column{3}{@{}>{\hspre}l<{\hspost}@{}}%
\column{E}{@{}>{\hspre}l<{\hspost}@{}}%
\>[3]{}\Keyword{data}\;\Conid{I}\;\mathrel{=}\;\Varid{i}{}\<[E]%
\ColumnHook
\end{hscode}\resethooks
(We have taken the liberty of introducing a lowercase constructor,
which would cause a syntax error in Haskell.)

Next, we have the following definition:

\begin{quote}
  \textbf{Definition:} A \textbf{complex number} is an expression of
  the form
\begin{hscode}\SaveRestoreHook
\column{B}{@{}>{\hspre}l<{\hspost}@{}}%
\column{4}{@{}>{\hspre}l<{\hspost}@{}}%
\column{E}{@{}>{\hspre}l<{\hspost}@{}}%
\>[4]{}\Varid{a}\;\Varid{+}\;\Varid{bi}\;\qquad \mathrm{or} \qquad\Varid{a}\;\Varid{+}\;\Varid{ib},\ {}\<[E]%
\ColumnHook
\end{hscode}\resethooks
  where \ensuremath{\Varid{a}} and \ensuremath{\Varid{b}} are real numbers, and \ensuremath{\Varid{i}} is the imaginary unit.
\end{quote}

This definition clearly points to the introduction of a syntax (notice
the keyword ``form'').   This is underlined by the presentation of
\emph{two} forms, which can suggest that the operation of
juxtaposing \ensuremath{\Varid{i}} (multiplication?) is not commutative.

A profitable way of dealing with such concrete syntax in functional
programming is to introduce an abstract representation of it in the
form of a datatype:
\begin{hscode}\SaveRestoreHook
\column{B}{@{}>{\hspre}l<{\hspost}@{}}%
\column{3}{@{}>{\hspre}l<{\hspost}@{}}%
\column{17}{@{}>{\hspre}l<{\hspost}@{}}%
\column{20}{@{}>{\hspre}l<{\hspost}@{}}%
\column{E}{@{}>{\hspre}l<{\hspost}@{}}%
\>[3]{}\Keyword{data}\;\Conid{Complex}\;{}\<[17]%
\>[17]{}\mathrel{=}\;{}\<[20]%
\>[20]{}\Conid{Plus}_1\;\mathbb{R}\;\mathbb{R}\;\Conid{I}\;{}\<[E]%
\\
\>[17]{}\mid \;{}\<[20]%
\>[20]{}\Conid{Plus}_2\;\mathbb{R}\;\Conid{I}\;\mathbb{R}{}\<[E]%
\ColumnHook
\end{hscode}\resethooks
We can give the translation from the abstract syntax to the concrete
syntax as a function \ensuremath{\Varid{show}}:
\begin{hscode}\SaveRestoreHook
\column{B}{@{}>{\hspre}l<{\hspost}@{}}%
\column{3}{@{}>{\hspre}l<{\hspost}@{}}%
\column{23}{@{}>{\hspre}l<{\hspost}@{}}%
\column{26}{@{}>{\hspre}l<{\hspost}@{}}%
\column{E}{@{}>{\hspre}l<{\hspost}@{}}%
\>[3]{}\Varid{show}\;{}\<[23]%
\>[23]{}\mathbin{:}\;{}\<[26]%
\>[26]{}\Conid{Complex}\;\to \;\Conid{String}{}\<[E]%
\\
\>[3]{}\Varid{show}\;(\Conid{Plus}_1\;\Varid{x}\;\Varid{y}\;\Varid{i})\;{}\<[23]%
\>[23]{}\mathrel{=}\;{}\<[26]%
\>[26]{}\Varid{show}\;\Varid{x}\;\plus \;\text{\tt \char34 ~+~\char34}\;\plus \;\Varid{show}\;\Varid{y}\;\plus \;\text{\tt \char34 i\char34}{}\<[E]%
\\
\>[3]{}\Varid{show}\;(\Conid{Plus}_2\;\Varid{x}\;\Varid{i}\;\Varid{y})\;{}\<[23]%
\>[23]{}\mathrel{=}\;{}\<[26]%
\>[26]{}\Varid{show}\;\Varid{x}\;\plus \;\text{\tt \char34 ~+~\char34}\;\plus \;\text{\tt \char34 i\char34}\;\plus \;\Varid{show}\;\Varid{y}{}\<[E]%
\ColumnHook
\end{hscode}\resethooks
The text continues with examples:

\begin{quote}
  For example, \ensuremath{\Varid{3}\;\Varid{+}\;\Varid{2}\;\Varid{i}}, \ensuremath{\frac{\Varid{7}}{\Varid{2}}\;\Varid{-}\;\frac{\Varid{2}}{\Varid{3}}\;\Varid{i}} , \ensuremath{\Varid{i}\;\pi\;\mathrel{=}\;\Varid{0}\;\Varid{+}\;\Varid{i}\;\pi} , and \ensuremath{\Varid{-3}\;\mathrel{=}\;\Varid{-3}\;\Varid{+}\;\Varid{0}\;\Varid{i}} are all complex numbers.  The last of these examples shows
  that every real number can be regarded as a complex number.
\end{quote}

The second example is somewhat problematic: it does not seem to be of
the form \ensuremath{\Varid{a}\;\Varid{+}\;\Varid{bi}}.  Given that the last two examples seem to introduce
shorthand for various complex numbers, let us assume that this one
does as well, and that \ensuremath{\Varid{a}\;\Varid{-}\;\Varid{bi}} can be understood as an abbreviation
of \ensuremath{\Varid{a}\;\Varid{+}\;(\Varid{-b})\;\Varid{i}}.

With this provision, in our notation the examples are written as
\ensuremath{\Conid{Plus}_1\;\Varid{3}\;\Varid{2}\;\Varid{i}}, \ensuremath{\Conid{Plus}_1\;\frac{\Varid{7}}{\Varid{2}}\;(\Varid{-}\;\frac{\Varid{2}}{\Varid{3}})\;\Varid{i}}, \ensuremath{\Conid{Plus}_2\;\Varid{0}\;\Varid{i}\;\pi}, \ensuremath{\Conid{Plus}_1\;(\Varid{-3})\;\Varid{0}\;\Varid{i}}.  We interpret the sentence ``The last of these examples
\ldots'' to mean that there is an embedding of the real numbers in
\ensuremath{\Conid{Complex}}, which we introduce explicitly:
\begin{hscode}\SaveRestoreHook
\column{B}{@{}>{\hspre}l<{\hspost}@{}}%
\column{3}{@{}>{\hspre}l<{\hspost}@{}}%
\column{E}{@{}>{\hspre}l<{\hspost}@{}}%
\>[3]{}\Varid{toComplex}\;\mathbin{:}\;\mathbb{R}\;\to \;\Conid{Complex}{}\<[E]%
\\
\>[3]{}\Varid{toComplex}\;\Varid{x}\;\mathrel{=}\;\Conid{Plus}_1\;\Varid{x}\;\Varid{0}\;\Varid{i}{}\<[E]%
\ColumnHook
\end{hscode}\resethooks
Again, at this stage there are many open questions.  For example, we
can assume that \ensuremath{\Varid{i1}} stands for the complex number \ensuremath{\Conid{Plus}_2\;\Varid{0}\;\Varid{i}\;\Varid{1}}, but
what about \ensuremath{\Varid{i}} by itself?  If juxtaposition is meant to denote some
sort of multiplication, then perhaps \ensuremath{\Varid{1}} can be considered as a unit,
in which case we would have that \ensuremath{\Varid{i}} abbreviates \ensuremath{\Varid{i1}} and therefore
\ensuremath{\Conid{Plus}_2\;\Varid{0}\;\Varid{i}\;\Varid{1}}.  But what about, say, \ensuremath{\Varid{2}\;\Varid{i}}?  Abbreviations with \ensuremath{\Varid{i}}
have only been introduced for the \ensuremath{\Varid{ib}} form, and not for the \ensuremath{\Varid{bi}} one!

The text then continues with a parenthetical remark which helps us
dispel these doubts:

\begin{quote}
  (We will normally use \ensuremath{\Varid{a}\;\Varid{+}\;\Varid{bi}} unless \ensuremath{\Varid{b}} is a complicated
  expression, in which case we will write \ensuremath{\Varid{a}\;\Varid{+}\;\Varid{ib}} instead. Either
  form is acceptable.)
\end{quote}

This remark suggests strongly that the two syntactic forms are meant to
denote the same elements, since otherwise it would be strange to say
``either form is acceptable''.  After all, they are acceptable by
definition.

Given that \ensuremath{\Varid{a}\;\Varid{+}\;\Varid{ib}} is only ``syntactic sugar'' for \ensuremath{\Varid{a}\;\Varid{+}\;\Varid{bi}}, we can
simplify our representation for the abstract syntax, eliminating one
of the constructors:
\begin{hscode}\SaveRestoreHook
\column{B}{@{}>{\hspre}l<{\hspost}@{}}%
\column{3}{@{}>{\hspre}l<{\hspost}@{}}%
\column{E}{@{}>{\hspre}l<{\hspost}@{}}%
\>[3]{}\Keyword{data}\;\Conid{Complex}\;\mathrel{=}\;\Conid{Plus}\;\mathbb{R}\;\mathbb{R}\;\Conid{I}{}\<[E]%
\ColumnHook
\end{hscode}\resethooks
In fact, since it doesn't look as though the type \ensuremath{\Conid{I}} will receive
more elements, we can dispense with it altogether:
\begin{hscode}\SaveRestoreHook
\column{B}{@{}>{\hspre}l<{\hspost}@{}}%
\column{3}{@{}>{\hspre}l<{\hspost}@{}}%
\column{E}{@{}>{\hspre}l<{\hspost}@{}}%
\>[3]{}\Keyword{data}\;\Conid{Complex}\;\mathrel{=}\;\Conid{PlusI}\;\mathbb{R}\;\mathbb{R}{}\<[E]%
\ColumnHook
\end{hscode}\resethooks
\noindent
(The renaming of the constructor from \ensuremath{\Conid{Plus}} to \ensuremath{\Conid{PlusI}} serves as a
guard against the case we have suppressed potentially semantically
relevant syntax.)

We read further:

\begin{quote}
  It is often convenient to represent a complex number by a single
  letter; \ensuremath{\Varid{w}} and \ensuremath{\Varid{z}} are frequently used for this purpose. If \ensuremath{\Varid{a}},
  \ensuremath{\Varid{b}}, \ensuremath{\Varid{x}}, and \ensuremath{\Varid{y}} are real numbers, and \ensuremath{\Varid{w}\;\mathrel{=}\;\Varid{a}\;\Varid{+}\;\Varid{bi}} and \ensuremath{\Varid{z}\;\mathrel{=}\;\Varid{x}\;\Varid{+}\;\Varid{yi}}, then we can refer to the complex numbers \ensuremath{\Varid{w}} and \ensuremath{\Varid{z}}. Note that
  \ensuremath{\Varid{w}\;\mathrel{=}\;\Varid{z}} if and only if \ensuremath{\Varid{a}\;\mathrel{=}\;\Varid{x}} and \ensuremath{\Varid{b}\;\mathrel{=}\;\Varid{y}}.
\end{quote}

First, let us notice that we are given an important semantic
information: \ensuremath{\Conid{PlusI}} is not just syntactically injective (as all
constructors are), but also semantically.  The equality on complex
numbers is what we would obtain in Haskell by using \ensuremath{\Varid{deriving}\;\Conid{Eq}}.

This shows that complex numbers are, in fact, isomorphic with pairs of
real numbers, a point which we can make explicit by re-formulating the
definition in terms of a type synonym:
\begin{hscode}\SaveRestoreHook
\column{B}{@{}>{\hspre}l<{\hspost}@{}}%
\column{3}{@{}>{\hspre}l<{\hspost}@{}}%
\column{E}{@{}>{\hspre}l<{\hspost}@{}}%
\>[3]{}\Keyword{newtype}\;\Conid{Complex}\;\mathrel{=}\;\Conid{C}\;(\mathbb{R},\ \mathbb{R}){}\<[E]%
\ColumnHook
\end{hscode}\resethooks
The point of the somewhat confusing discussion of using ``letters'' to
stand for complex numbers is to introduce a substitute for \emph{pattern
  matching}, as in the following definition:

\begin{quote}
  \textbf{Definition:} If \ensuremath{\Varid{z}\;\mathrel{=}\;\Varid{x}\;\Varid{+}\;\Varid{yi}} is a complex number (where \ensuremath{\Varid{x}}
  and \ensuremath{\Varid{y}} are real), we call \ensuremath{\Varid{x}} the \textbf{real part} of \ensuremath{\Varid{z}} and denote it
  \ensuremath{\Conid{Re}\;(\Varid{z})}. We call \ensuremath{\Varid{y}} the \textbf{imaginary part} of \ensuremath{\Varid{z}} and denote it \ensuremath{\Conid{Im}\;(\Varid{z})}:
\begin{hscode}\SaveRestoreHook
\column{B}{@{}>{\hspre}l<{\hspost}@{}}%
\column{3}{@{}>{\hspre}l<{\hspost}@{}}%
\column{10}{@{}>{\hspre}l<{\hspost}@{}}%
\column{13}{@{}>{\hspre}l<{\hspost}@{}}%
\column{26}{@{}>{\hspre}l<{\hspost}@{}}%
\column{29}{@{}>{\hspre}l<{\hspost}@{}}%
\column{E}{@{}>{\hspre}l<{\hspost}@{}}%
\>[3]{}\Conid{Re}\;(\Varid{z})\;{}\<[10]%
\>[10]{}\mathrel{=}\;{}\<[13]%
\>[13]{}\Conid{Re}\;(\Varid{x}\;\Varid{+}\;\Varid{yi})\;{}\<[26]%
\>[26]{}\mathrel{=}\;{}\<[29]%
\>[29]{}\Varid{x}{}\<[E]%
\\
\>[3]{}\Conid{Im}\;(\Varid{z})\;{}\<[10]%
\>[10]{}\mathrel{=}\;{}\<[13]%
\>[13]{}\Conid{Im}\;(\Varid{x}\;\Varid{+}\;\Varid{yi})\;{}\<[26]%
\>[26]{}\mathrel{=}\;{}\<[29]%
\>[29]{}\Varid{y}{}\<[E]%
\ColumnHook
\end{hscode}\resethooks
\end{quote}

This is rather similar to Haskell's \emph{as-patterns}:
\begin{hscode}\SaveRestoreHook
\column{B}{@{}>{\hspre}l<{\hspost}@{}}%
\column{3}{@{}>{\hspre}l<{\hspost}@{}}%
\column{22}{@{}>{\hspre}l<{\hspost}@{}}%
\column{26}{@{}>{\hspre}l<{\hspost}@{}}%
\column{E}{@{}>{\hspre}l<{\hspost}@{}}%
\>[3]{}\Conid{Re}\;\mathbin{:}\;\Conid{Complex}\;{}\<[22]%
\>[22]{}\to \;{}\<[26]%
\>[26]{}\mathbb{R}{}\<[E]%
\\
\>[3]{}\Conid{Re}\;\Varid{z}\;\!@\!\;(\Conid{C}\;(\Varid{x},\ \Varid{y}))\;{}\<[22]%
\>[22]{}\mathrel{=}\;{}\<[26]%
\>[26]{}\Varid{x}{}\<[E]%
\ColumnHook
\end{hscode}\resethooks
\begin{hscode}\SaveRestoreHook
\column{B}{@{}>{\hspre}l<{\hspost}@{}}%
\column{3}{@{}>{\hspre}l<{\hspost}@{}}%
\column{22}{@{}>{\hspre}l<{\hspost}@{}}%
\column{26}{@{}>{\hspre}l<{\hspost}@{}}%
\column{E}{@{}>{\hspre}l<{\hspost}@{}}%
\>[3]{}\Conid{Im}\;\mathbin{:}\;\Conid{Complex}\;{}\<[22]%
\>[22]{}\to \;{}\<[26]%
\>[26]{}\mathbb{R}{}\<[E]%
\\
\>[3]{}\Conid{Im}\;\Varid{z}\;\!@\!\;(\Conid{C}\;(\Varid{x},\ \Varid{y}))\;{}\<[22]%
\>[22]{}\mathrel{=}\;{}\<[26]%
\>[26]{}\Varid{y}{}\<[E]%
\ColumnHook
\end{hscode}\resethooks
\noindent
a potential source of confusion being that the symbol \ensuremath{\Varid{z}} introduced
by the as-pattern is not actually used on the right-hand side of the
equations.

The use of as-patterns such as ``\ensuremath{\Varid{z}\;\mathrel{=}\;\Varid{x}\;\Varid{+}\;\Varid{yi}}'' is repeated throughout
the text, for example in the definition of the algebraic operations on
complex numbers:

\begin{quote}
  \textbf{The sum and difference of complex numbers}

  If \ensuremath{\Varid{w}\;\mathrel{=}\;\Varid{a}\;\Varid{+}\;\Varid{bi}} and \ensuremath{\Varid{z}\;\mathrel{=}\;\Varid{x}\;\Varid{+}\;\Varid{yi}}, where \ensuremath{\Varid{a}}, \ensuremath{\Varid{b}}, \ensuremath{\Varid{x}}, and \ensuremath{\Varid{y}} are real numbers,
  then
\begin{hscode}\SaveRestoreHook
\column{B}{@{}>{\hspre}l<{\hspost}@{}}%
\column{3}{@{}>{\hspre}l<{\hspost}@{}}%
\column{6}{@{}>{\hspre}l<{\hspost}@{}}%
\column{9}{@{}>{\hspre}l<{\hspost}@{}}%
\column{12}{@{}>{\hspre}l<{\hspost}@{}}%
\column{15}{@{}>{\hspre}l<{\hspost}@{}}%
\column{24}{@{}>{\hspre}l<{\hspost}@{}}%
\column{27}{@{}>{\hspre}l<{\hspost}@{}}%
\column{E}{@{}>{\hspre}l<{\hspost}@{}}%
\>[3]{}\Varid{w}\;{}\<[6]%
\>[6]{}\Varid{+}\;{}\<[9]%
\>[9]{}\Varid{z}\;{}\<[12]%
\>[12]{}\mathrel{=}\;{}\<[15]%
\>[15]{}(\Varid{a}\;\Varid{+}\;\Varid{x})\;{}\<[24]%
\>[24]{}\Varid{+}\;{}\<[27]%
\>[27]{}(\Varid{b}\;\Varid{+}\;\Varid{y})\;\Varid{i}{}\<[E]%
\\[\blanklineskip]%
\>[3]{}\Varid{w}\;{}\<[6]%
\>[6]{}\Varid{-}\;{}\<[9]%
\>[9]{}\Varid{z}\;{}\<[12]%
\>[12]{}\mathrel{=}\;{}\<[15]%
\>[15]{}(\Varid{a}\;\Varid{-}\;\Varid{x})\;{}\<[24]%
\>[24]{}\Varid{+}\;{}\<[27]%
\>[27]{}(\Varid{b}\;\Varid{-}\;\Varid{y})\;\Varid{i}{}\<[E]%
\ColumnHook
\end{hscode}\resethooks
\end{quote}

With the introduction of algebraic operations, the language of complex
numbers becomes much richer.  We can describe these operations in a
\emph{shallow embedding} in terms of the concrete datatype \ensuremath{\Conid{Complex}}, for
example:
\begin{hscode}\SaveRestoreHook
\column{B}{@{}>{\hspre}l<{\hspost}@{}}%
\column{3}{@{}>{\hspre}l<{\hspost}@{}}%
\column{8}{@{}>{\hspre}l<{\hspost}@{}}%
\column{11}{@{}>{\hspre}l<{\hspost}@{}}%
\column{28}{@{}>{\hspre}l<{\hspost}@{}}%
\column{31}{@{}>{\hspre}l<{\hspost}@{}}%
\column{E}{@{}>{\hspre}l<{\hspost}@{}}%
\>[3]{}(\Varid{+})\;{}\<[8]%
\>[8]{}\mathbin{:}\;{}\<[11]%
\>[11]{}\Conid{Complex}\;\to \;\Conid{Complex}\;\to \;\Conid{Complex}{}\<[E]%
\\
\>[3]{}(\Conid{C}\;(\Varid{a},\ \Varid{b}))\;\Varid{+}\;(\Conid{C}\;(\Varid{x},\ \Varid{y}))\;{}\<[28]%
\>[28]{}\mathrel{=}\;{}\<[31]%
\>[31]{}\Conid{C}\;((\Varid{a}\;\Varid{+}\;\Varid{x}),\ (\Varid{b}\;\Varid{+}\;\Varid{y})){}\<[E]%
\ColumnHook
\end{hscode}\resethooks
\noindent
or we can build a datatype of ``syntactic'' Complex numbers from the
algebraic operations to arrive at a \emph{deep embedding}:
\begin{hscode}\SaveRestoreHook
\column{B}{@{}>{\hspre}l<{\hspost}@{}}%
\column{3}{@{}>{\hspre}l<{\hspost}@{}}%
\column{23}{@{}>{\hspre}l<{\hspost}@{}}%
\column{26}{@{}>{\hspre}l<{\hspost}@{}}%
\column{33}{@{}>{\hspre}l<{\hspost}@{}}%
\column{48}{@{}>{\hspre}l<{\hspost}@{}}%
\column{E}{@{}>{\hspre}l<{\hspost}@{}}%
\>[3]{}\Keyword{data}\;\Conid{ComplexSyntax}\;{}\<[23]%
\>[23]{}\mathrel{=}\;{}\<[26]%
\>[26]{}\Varid{i}\;{}\<[E]%
\\
\>[23]{}\mid \;{}\<[26]%
\>[26]{}\Conid{ToComplex}\;\mathbb{R}\;{}\<[E]%
\\
\>[23]{}\mid \;{}\<[26]%
\>[26]{}\Conid{Plus}\;{}\<[33]%
\>[33]{}\Conid{ComplexSyntax}\;{}\<[48]%
\>[48]{}\Conid{ComplexSyntax}\;{}\<[E]%
\\
\>[23]{}\mid \;{}\<[26]%
\>[26]{}\Conid{Times}\;{}\<[33]%
\>[33]{}\Conid{ComplexSyntax}\;{}\<[48]%
\>[48]{}\Conid{ComplexSyntax}\;{}\<[E]%
\\
\>[23]{}\mid \;{}\<[26]%
\>[26]{}\Varid{...}{}\<[E]%
\ColumnHook
\end{hscode}\resethooks
The type \ensuremath{\Conid{ComplexSyntax}} can then be turned into an abstract datatype,
by hiding the representation and providing corresponding operations
like \ensuremath{(\Varid{+})\;\mathrel{=}\;\Conid{Plus}}, etc.
Deep embedding offers a cleaner separation between syntax and
semantics, making it possible to compare and factor out the common
parts of various languages.  For the computer science students, this
is a way of approaching structural algebra; for the mathematics
students, this is a way to learn the ideas of abstract datatypes, type
classes, folds, by relating them to the familiar notions of
mathematical structures and homomorphisms (see
\cite{gibbons2014folding} for a discussion of the relationships
between deep and shallow embeddings and folds).
We want to show the students both the shallow and the deep approach
and help them understand when more or less focus on syntax is helpful.

Adams and Essex then proceed to introduce the geometric
interpretation of complex numbers, i.e., the isomorphism between
complex numbers and points in the Euclidean plane as pairs of
coordinates.  The isomorphism is not given a name, but we can use the
constructor \ensuremath{\Conid{C}} defined above.  They then define the polar
representation of complex numbers, in terms of modulus and argument:

\begin{quote}
  The distance from the origin to the point \ensuremath{(\Varid{a},\ \Varid{b})} corresponding to
  the complex number \ensuremath{\Varid{w}\;\mathrel{=}\;\Varid{a}\;\Varid{+}\;\Varid{bi}} is called the \textbf{modulus} of \ensuremath{\Varid{w}} and is
  denoted by \ensuremath{\lvert{}\Varid{w}\rvert{}} or \ensuremath{\lvert{}\Varid{a}\;\Varid{+}\;\Varid{bi}\rvert{}}:
\begin{hscode}\SaveRestoreHook
\column{B}{@{}>{\hspre}l<{\hspost}@{}}%
\column{3}{@{}>{\hspre}l<{\hspost}@{}}%
\column{E}{@{}>{\hspre}l<{\hspost}@{}}%
\>[3]{}\lvert{}\Varid{w}\rvert{}\;\mathrel{=}\;\lvert{}\Varid{a}\;\Varid{+}\;\Varid{bi}\rvert{}\;\mathrel{=}\;\sqrt{\Varid{a}^2 + \Varid{b}^2}{}\<[E]%
\ColumnHook
\end{hscode}\resethooks
  If the line from the origin to \ensuremath{(\Varid{a},\ \Varid{b})} makes angle \ensuremath{\theta} with the
  positive direction of the real axis (with positive angles measured
  counterclockwise), then we call \ensuremath{\theta} an \textbf{argument} of the
  complex number \ensuremath{\Varid{w}\;\mathrel{=}\;\Varid{a}\;\Varid{+}\;\Varid{bi}} and denote it by \ensuremath{\Varid{arg}\;(\Varid{w})} or \ensuremath{\Varid{arg}\;(\Varid{a}\;\Varid{+}\;\Varid{bi})}.

\end{quote}

Here, the constant repetitions of ``\ensuremath{\Varid{w}\;\mathrel{=}\;\Varid{a}\;\Varid{+}\;\Varid{bi}}'' and ``\ensuremath{\Varid{f}\;(\Varid{w})} or \ensuremath{\Varid{f}\;(\Varid{a}\;\Varid{+}\;\Varid{bi})}'' are caused not just by the unavailability of
pattern-matching, but also by the absence of the explicit isomorphism
\ensuremath{\Conid{C}}.  We need only use \ensuremath{\lvert{}\Conid{C}\;(\Varid{a},\ \Varid{b})\rvert{}\;\mathrel{=}\;\sqrt{\Varid{a}^2 + \Varid{b}^2}}, making clear
that the modulus and arguments are actually defined by pattern
matching.

Once the principal argument has been defined as the unique argument in
the interval \ensuremath{(\Varid{-}\;\pi,\ \pi]}, the way is opened to a different
interpretation of complex numbers (usually called the \emph{polar
  representation} of complex numbers):
\begin{hscode}\SaveRestoreHook
\column{B}{@{}>{\hspre}l<{\hspost}@{}}%
\column{3}{@{}>{\hspre}l<{\hspost}@{}}%
\column{E}{@{}>{\hspre}l<{\hspost}@{}}%
\>[3]{}\Keyword{newtype}\;\Conid{Complex'}\;\mathrel{=}\;\Conid{C'}\;(\mathbb{R}_{\ge 0},\ (\Varid{-}\;\pi,\ \pi]){}\<[E]%
\ColumnHook
\end{hscode}\resethooks
\ensuremath{\Conid{C'}} constructs a ``geometric'' complex number from a non-negative
modulus and a principal argument; the (non-implementable) constraints
on the types ensure uniqueness of representation.

The importance of this alternative representation is that the
operations on its elements have a different natural interpretation,
namely as geometrical operations.  For example, multiplication with
\ensuremath{\Conid{C'}\;(\Varid{m},\ \theta)} represents a re-scaling of the Euclidean plane with a
factor \ensuremath{\Varid{m}}, coupled with a rotation with angle \ensuremath{\theta}.  Thus,
multiplication with \ensuremath{\Varid{i}} (which is \ensuremath{\Conid{C'}\;(\Varid{1},\ \frac{\pi}{\Varid{2}})} in polar
representation) results in a counterclockwise rotation of the plane by
90°.  This interpretation of \ensuremath{\Varid{i}} seems independent of the originally
proposed arithmetical one (``the square root of -1''), and the polar
representation of complex numbers leads to a different, geometrical
language.

It can be an interesting exercise to develop this language (of
scalings, rotations, etc.) ``from scratch'', without reference to
complex numbers.  In a deep embedding, the result is a datatype
representing a syntax that is quite different from the one suggested
by the algebraic operations.  The fact that this language can also be
given semantics in terms of complex numbers could then be seen as
somewhat surprising, and certainly in need of proof.  This would
introduce in a simple setting the fact that many fundamental theorems
in mathematics establish that two languages with different syntaxes
have, in fact, the same semantics.  A more elaborate example is that
of the identity of the language of matrix manipulations as implemented
in Matlab and that of linear transformations.  At the undergraduate
level, the most striking example is perhaps that of the identity of
holomorphic functions (the language of complex derivatives) and (regular)
analytic functions (the language of complex power series).

\section{Conclusions and future work}

We have presented the basic ingredients of an approach that uses
functional programming as a way of helping students deal with
classical mathematics and its applications:

\begin{itemize}
\item make functions and the types explicit

\item use types as carriers of semantic information, not just variable
  names

\item introduce functions and types for implicit operations such as
  the power series interpretation of a sequence

\item use a calculational style for proofs

\item organise the types and functions in DSLs
\end{itemize}

Given the main course objective, enabling the students to better
tackle mathematical domains by applying the computing science
perspective, we intend to measure how well the students do in ulterior
courses that require mathematical competence.  For example, we will
measure the percentage of students who, having taken DSLsofMath, pass the
third-year courses
\emph{\href{https://www.student.chalmers.se/sp/course?course_id=21865}{Transforms,
    signals and systems}} and
\emph{\href{https://www.student.chalmers.se/sp/course?course_id=21303}{Control
    Theory (Reglerteknik)}}, which are current major stumbling blocks.
Since the course will, at least initially, be an elective one, we will
also have the possibility of comparing the results with those of a
control group (students who have not taken the course).

The lessons in this course will be organised around the active reading
of mathematical texts (suitably prepared in advance).  In the opening
lessons, we will deal with domains of mathematics which are relatively
close to functional programming, such as elementary category theory,
in order to have the chance to introduce newcomers to functional
programming, and the students in general to our approach.

After that, the selection of the subjects will mostly be dictated by
the requirements of the engineering curriculum.  They will contain:

\begin{itemize}
\item basic properties of complex numbers

\item the exponential function

\item elementary functions

\item holomorphic functions

\item the Laplace transform

\end{itemize}

We shall take advantage of the fact that some parts of these topics
have been treated before from a functional programming perspective
\cite{mcilroy1999functional, mcilroy2001music,
  pavlovic1999coalgebra}.

One of the important course elements we have left out of this paper is
that of using the modelling effort performed in the course for the
production of actual mathematical software.  One of the reasons for
this omission is that we wanted to concentrate on the more conceptual
part that corresponds to the specification of that software, and as
such is a prerequisite for it.  The development of implementations on
the basis of these specifications will be the topic of most of the
exercise sessions we will organise.  That the computational
representation of mathematical concepts can greatly help with their
understanding was conclusively shown by Sussman and Wisdom in their
recent book on differential geometry \cite{sussman2013functional}.

On the other hand, classical mathematical theorems often lead to
non-implementable specifications (for example, there is no algorithm
for finding the minima and maxima of arbitrary continuous functions on
a closed interval, although we have an easy classical proof of their
existence).  There are many possibilities of dealing with such cases,
and we shall explore some of them in the exercises sessions.  For
instance,
in scientific programming, one is often interested in correctness ``up
to implication'': the program would work as expected, say, if one
would use real numbers instead of floating-point values.
Such counterfactuals are impossible to test but they can be encoded as
types and proven~\cite{ionescu2013testing}.
%
%

We believe that this approach can offer an introduction to computer
science for the mathematics students.  We plan to actively involve the
mathematics faculty at Chalmers, via guest lectures and regular
meetings, in order to find the suitable middle ground we alluded to in
the introduction: between a presentation that is too explicit, turning
the student into a spectator of endless details, and one that is too
implicit and leaves so much for the students to do that they are
overwhelmed.  Ideally, some of the features of our approach would be
worked into the earlier mathematical courses.

The computer science perspective has been quite successful in
influencing the presentation of discrete mathematics.  For example,
the classical textbook of Gries and Schneider, \emph{A Logical
  Approach to Discrete Math} \cite{gries1993logical}, has been
well-received by both computer scientists and mathematicians.  When it
comes to continuous mathematics, however, there is no such influence
to be felt.  The work presented here represents the starting point of
an attempt to change this state of affairs.

\bibliographystyle{eptcs}
\bibliography{dslm}




\end{document}